# Electromagnetic Field Exposure Assessment and Mitigation Strategies for Wireless Power Transfer Systems: A Review and Future Perspectives

*Akimasa Hirata, Teruo Onishi, Naoki Shinohara, Valerio De Santis, Yinliang Diao, Mauro Feliziani, Takashi Hikage, Junqing Lan, Francesca Maradei, Keishi Miwa, Alexander Prokop, Yujun Shin, Seungyoung Ahn*

*Abstract*—Wireless power transfer (WPT) technologies are increasingly being applied in fields ranging from consumer electronics and electric vehicles to space-based energy systems and medical implants. While WPT offers contactless power delivery, it introduces electromagnetic field (EMF) emissions, necessitating careful assessment to address safety and public health concerns. Exposure guidelines developed by ICNIRP and IEEE define frequency-dependent limits based on internal quantities, such as electric field strength and specific absorption rate, intended to prevent tissue nerve stimulation < 100 kHz and heating > 100 kHz, respectively. Complementing these guidelines, assessment standards including the International Electrotechnical Commission (IEC)/IEEE 63184 and IEC Technical Report 63377, provide practical procedures for evaluating the EMF exposure in WPT systems. This review offers a comparative overview of major WPT modalities, with a focus on recent developments in computational dosimetry and standardized assessment techniques for the complex, non-uniform fields typical of WPT environments. It also discusses electromagnetic interference with medical devices and exposure scenarios involving partial body proximity and various postures. A notable observation across modalities is the considerable variability, often spanning an order of magnitude, in the allowable transfer power, depending on the field distribution and assessment approach. Remaining challenges include the lack of harmonized guidance for intermediate frequencies and localized exposure, underscoring the importance of further coordination in international standardization efforts. Addressing these issues is essential for the safe and widespread deployment of WPT technologies.

*Index Terms*— assessment methods; dosimetry; exposure guidelines; mitigation; standardization

## I. INTRODUCTION

WIRELESS power transfer (WPT) has emerged as a pivotal technology for contactless energy delivery across diverse applications, including consumer electronics, electric vehicles (EVs), medical implants, and space systems [1-5]. Summaries of WPT applications have been published by the International Telecommunication Union–Radiocommunication Sector (ITU–R) [6-8]. By eliminating the need for physical connectors, WPT enhances usability, enables sealed or rotating mechanisms, and broadens the feasibility of fully autonomous devices. However, unlike conventional wired power systems, WPT inherently emits electromagnetic fields (EMFs) into open environments, potentially causing unintended human exposure or higher field strengths than those of wireless communication systems, even when spatially confined. As WPT continues to scale in power, transmission distance, and deployment, EMF protection and exposure assessments have become essential for societal acceptance and regulatory development [9, 10].

Various WPT technologies employ different coupling mechanisms, such as radiative [11], inductive [12], or resonant [1], leading to distinct EMF characteristics. Microwave power transmission (MPT) [13, 14] employs focused far-field radiation, typically at 920 MHz, 2.45 GHz, or 5.8 GHz, extending to millimeter-wave frequencies [13], for long-distance energy transmission. The ITU–R recommends frequencies suitable for far-field WPT [15]. In contrast, near-field techniques, such as inductive or resonant coupling, operate at lower frequencies (typically below 13.56 MHz) [1], generating spatially confined magnetic fields for short- to mid-range power delivery. The ITU–R also recommends suitable frequencies for near-field WPT [16]. These differences lead to distinct electromagnetic exposure profiles, affecting field intensity, spatial distribution, and potential interactions with the human body.

TABLE I. SUMMARY OF FUNDAMENTAL CHARACTERISTICS AND EMF CHARACTERISTICS FOR DIFFERENT WPT SYSTEMS

| WPT Types | Coupling Mechanism | Frequency Band | Efficiency | Distance | Power Range | EMF Characteristics / Safety Notes |
|---|---|---|---|---|---|---|
| Microwave Power Transmission | EM radiation (far-field) | 2.45 GHz 5.8 GHz 24–60 GHz | Low–High | Long-range (10 m–km) | High (W–kW) | Directional far-field beam; narrow main lobe; exposure requires beam control |
| Inductive Coupling | Magnetic field (near-field) | 100 kHz– 13.56 MHz | High (>90%) | Short-range (<10 cm) | High (W–kW) | Localized reactive magnetic field; rapid decay; low stray exposure risk |
| Resonant Coupling | Magnetic/ EM resonance (mid-field) | 85 kHz– 10 MHz | High (~80%–90%) | Mid-range (10 cm–2 m) | Medium–High | Non-radiative oscillating field; wider spatial spread; $k$ and $Q$ determine field leakage |

Despite the growing interest in WPT, most existing review articles focus primarily on circuit design, energy efficiency, power management, and integration with communication systems, such as simultaneous wireless information and power transfer [17, 18]. However, few reviews have addressed human protection issues across different WPT modalities [19], and those have often adopted a limited product-assessment perspective [20]. Moreover, the wide variation in spatial EMF characteristics, from narrow-beam lobes in MPT to confined reactive fields in inductive systems, requires tailored modeling and assessment approaches. This is critical for evaluating compliance with international exposure guidelines, such as those established by the ICNIRP [21] and IEEE [22], which set the specific absorption rate (SAR) within the body and frequency-dependent permissible field strength. A thorough approach to compliance assessment, encompassing both human exposure and product compliance, is essential for future WPT systems. Concurrently, mitigation strategies aimed at reducing human exposure through system-level design are becoming increasingly critical, particularly in high-power or proximity-based applications.

This review addresses this gap by presenting a comparative, human protection-oriented overview of three representative WPT technologies: MPT, inductive coupling, and resonant coupling. For each of them, the underlying physical principles, frequency bands, coupling mechanisms, and transmission ranges are described in detail. Particular emphasis was placed on the spatial distribution of EMFs, their decay characteristics, and their implications for human exposure under realistic conditions. Furthermore, because some WPT systems generate strong and spatially complex EMFs, mitigation strategies for reducing or controlling human exposure are increasingly important. In addition, representative mitigation strategies for each method are discussed, offering practical insights into reducing or controlling EMF exposure through system-level design. By emphasizing field characteristics and mitigation approaches, this review aims to provide a foundational understanding of EMF exposure implications and support the future development of safe, effective WPT systems.

## II. REVIEW OF WPT SYSTEMS

WPT systems are broadly categorized into three types based on their coupling mechanisms, operating frequency bands, and transmission characteristics: MPT, inductive coupling, and resonant coupling. Table I summarizes the key features of these systems, including coupling type, frequency range, power levels, and EMF characteristics relevant to safety considerations. This overview forms the foundation for the detailed technical descriptions and exposure safety implications reviewed in Sections IV.A.–IV.C.

### A. Microwave Power Transmission

MPT refers to the long-range delivery of electrical energy via focused electromagnetic waves, typically operating in microwave frequency bands, such as the industrial, scientific, and medical (ISM) bands (2.45 and 5.8 GHz) and higher frequencies (for example, 24 and 60 GHz). This concept was first demonstrated by W. C. Brown in the 1960s [23], who converted power into microwaves, transmitted it via highly directional antennas, and converted it back to electricity using rectennas [5, 24]. Recently, this method has been termed "narrow-beam WPT." This method is central to proposals for space-based solar power systems that collect solar energy in orbit and transmit it to ground stations. Field tests in countries such as Japan [13, 14], the United States [13], the United Kingdom [25], and China [14, 26] have demonstrated WPT over distances spanning tens to hundreds of kilometers. The core components include phased-array antennas that allow dynamic beam steering and high-efficiency rectennas tailored to specific frequency bands [27]. MPT systems produce sharply confined far-field radiation patterns, with field strength concentrated in the main beam lobe and intensity decreasing rapidly outside this region. Beam width depends on the array size and frequency, with higher frequencies enabling tighter spatial confinement. In contrast, the narrow-beam WPT has not been adopted in the emerging MPT market of the 21st century. The commercial WPT system is named "wide-beam WPT," which distributes low-power radio waves broadly to multiple users, similar to radio-frequency identification (RFID) or mobile devices. The Japanese government has established new radio regulations for far-field WPT in May 2022, specifying systems at 920 MHz-1 W, 2.4 GHz-15 W, and 5.7 GHz-32 W [3] in consideration of the coexistence of existing radio applications and human safety issues. As of July 2025, over 640 WPT base stations operating at 920 MHz have been installed in buildings across Japan, and the WPT business has expanded within Japan and internationally.

### B. Inductive Coupling

Inductive coupling facilitates power transfer over short distances, generally limited to a few centimeters, by leveraging the magnetic interaction between a pair of coils. One coil generates a time-varying magnetic field that induces a current in the nearby receiving coil. Systems of this type typically operate in the 100 kHz–13.56 MHz range, where standardized ISM bands (such as 6.78 and 13.56 MHz) are commonly used. Numerous impact studies on in-band and close-band radio systems have been conducted in Japan to determine coexisting conditions. Applications span household devices, such as electric toothbrushes, wireless charging pads, and certain medical implants. The method is highly efficient under tight alignment and proximity, and its relatively simple architecture ensures cost-effectiveness and reliability.

The magnetic field generated by the transmitting coil is highly localized, exhibiting steep spatial attenuation—often proportional to the inverse cube of the distance—which results in strong coupling only within a narrow zone near the coil. Beyond this region, field strength and induced power decrease rapidly.

## C. Resonant Coupling

Resonant coupling enhances the efficiency and range of inductive systems by tuning the transmitting and receiving elements to a shared resonant frequency, typically between 100 kHz and 10 MHz. In addition, non-ISM 85 kHz is also permitted for EV wireless chargers worldwide [16]. This approach gained considerable attention following a 2006 MIT-led experiment that demonstrated mid-range energy transfer using magnetic resonance. Operational efficiency depends on the interaction between the coupling coefficient ($k$) and quality factor ($Q$) of the resonator. Well-designed high-$Q$ systems enable substantial energy transfer over distances exceeding those of basic inductive coupling. Typical applications include the wireless charging of EVs and other medium-range consumer electronics.

Resonant systems produce strong magnetic fields that are broader and more spatially distributed than those of simple inductive systems. Although these fields remain nonradiative, they decay more gradually with distance, particularly in high-$Q$ designs. The field concentrates around the resonant loops and extends up to one or two meters, depending on coil geometry and operating frequency.

## D. Comparison and Practical Considerations for Near-Field Coupling with Biological Tissues

Compared to inductive coupling, magnetic resonant coupling WPT systems are suitable for mid-range power transfer and offer greater transmitter/receiver flexibility. Although magnetic resonant coupling primarily transfers energy between resonators with minimal interaction with nonresonant objects, adjacent objects (metallic or ferrimagnetic materials) or the human body can alter the impedance of the resonant coil or perturb the magnetic flux path. This may change the resonant frequency and degrade the resonator quality factor and coupling coefficient, thereby reducing the transfer efficiency. Furthermore, the induced electric fields pose potential health risks. Inductive coupling is commonly used for short-distance (<40 mm) power transfer applications, with power levels ranging from a few watts to tens of kilowatts. It is widely applied in sensors and implantable devices owing to its high transfer efficiency and ease of implementation. However, its efficiency declines rapidly as coil separation increases.

## III.   FRAMEWORK OF HUMAN PROTECTION: GUIDELINES AND ASSESSMENT

## A. Exposure Guidelines

The World Health Organization (WHO) references two international guidelines and standards for human exposure: the ICNIRP guidelines [21, 28] and the IEEE C95.1 standard [22]. The ICNIRP operates as an independent nongovernmental body in formal relations with the WHO, while the IEEE C95.1 standard is developed through an open consensus process within its Standards Coordinating Committee. These guidelines and standards provide a framework for protecting against potential adverse health effects. Although developed independently, they share similar goals of human protection.

Both bodies adopt an evidence-based approach, requiring substantiated physical plausibility and consistency across multiple studies before health effects are considered in guideline development. The lowest threshold for adverse health effects is the nerve stimulation effect at frequencies < 100 kHz and heating > 100 kHz. Exposure at frequencies > 100 kHz is further divided into local and core temperature rises for local and whole-body exposures, respectively.

To derive these limits, thresholds for adverse health effects were identified, primarily from human volunteer studies, with sufficient margins applied. The threshold for nervous stimulation depends on frequency, and the ICNIRP guidelines [21] set limits at 5°C for local temperature increases and 1°C for core temperature increases. Specifically, to account for interindividual variability, uncertainties in biological data, and other factors, reduction factors [29]—also referred to as safety factors [22]—were applied in deriving the exposure limits.

The guidelines distinguish between basic restrictions and reference levels (Fig. 1). Basic restrictions refer to limits on internal electrical quantities directly linked to adverse effects. These include the induced electric field strength (V/m) at frequencies < 100 kHz, SAR (W/kg) from 100 kHz to 6 GHz, and absorbed power density (APD) > 6 GHz. For low-frequency (LF) exposures, following ICNIRP guidelines [28], the internal (induced) electric field averaged over a 2-mm cubic volume was employed, whereas a 5-mm line average value was utilized in the IEEE C95.1 standard [22]. These physical quantities are well correlated with adverse health effects: the internal electric field is the physical agent responsible for nerve stimulation; SAR averaged over 10 g of tissue and APD averaged over 4 cm² are correlated with tissue (internal) temperature and body surface temperature rise, respectively [30, 31]. In addition, the whole-body averaged SAR is used for whole-body (uniform field) exposures to correlate with core temperature rise [32].

Reference levels are permissible field strengths, expressed in terms of external physical quantities, such as the magnetic field strength (A/m), electric field strength (V/m), and power density (W/m²) in free space. These are intended as conservative levels based on the assumption of uniform field exposure conditions. Compliance with reference levels generally ensures compliance with basic restrictions. However, direct dosimetric evaluation is required in complex exposure scenarios, such as those involving WPT systems.

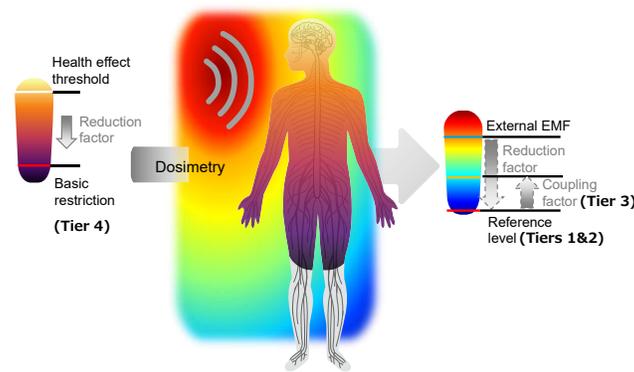

Fig. 1. Illustration of the process for deriving restrictions in international guidelines. Thresholds for substantiated adverse health effects are first identified. Basic restrictions (BR) are then set by applying appropriate safety or reduction factors. Finally, reference levels (RL) are determined using computational dosimetry with conservative assumptions. "Tier" in the figure corresponds to the definition in assessment standard (Sec. III D).

## B. Computational Aspects and Their Application to Exposure Assessment

Exposure guidelines use anatomically realistic human models to identify electrical quantities induced in the body that are linked to adverse health effects and the derivation of exposure limits. These human body or partial-body models have a spatial resolution between a fraction of a millimeter and a few millimeters, sufficient to represent the induced physical quantities.

The field distribution does not change substantially over a few millimeters, and a comparable value is obtained across different metrics [33]. In addition, the ICNIRP guidelines [28] use the 99th percentile of the tissue to exclude computational artifacts in induced electric fields. Although this metric is useful for exposure to a uniform field, the induced electric field is underestimated for local exposure [34]. Moreover, a human body model with a resolution of a few millimeters may be insufficient for certain body parts, such as the armpit, where skin-to-skin contact [22] can lead to unusual current flow [35]. Therefore, certain post-processing should be considered when deriving the limit and when applying it to conformity assessment. Several computational approaches have been proposed to eliminate these weaknesses [36, 37].

For RF (radio-frequency) exposures over 100 kHz, SAR averaged over 10 g of tissues and the whole body was used as a metric to relate to local and core temperature rises. For the former, the side length of the cube was approximately 22 mm for a volume of 10 g (assuming a tissue density of 1000 kg/m$^3$) [38]. The detailed computational procedure is provided in a product-specific conformity standard [39].

For frequencies above 6 GHz, the APD averaged over 4 cm$^2$ was used as a metric to relate to local surface temperature rise owing to the shallow penetration depth at these higher frequencies. An additional averaging area of 1 cm$^2$ is considered for frequencies above 30 GHz for narrow-beam radiation. Numerical methods for APD above 6 GHz for conformity assessment are currently under development [40].

Computational studies have shown that the 10 g average SAR correlates best with peak local temperature rise below 3 GHz, whereas APD more accurately predicts skin surface temperature rise above 10 GHz. The maximum heating factor for the averaged SAR was approximately 0.25°C/W·kg for skin tissue and 0.1°C/W·kg for brain tissue. The heating factor for the APD averaged over 4 cm$^2$ was 0.025°C/W·m$^2$ [41]. The transition frequency from SAR to APD was harmonized to 6 GHz between the latest versions of the ICNIRP guidelines and IEEE standards. For whole-body exposure, dosimetry shows that an average whole-body SAR of approximately 4–6 W/kg is required to increase the body core temperature by 1°C, depending on the sweating rate and mass-to-surface area ratio [42].

The voxel size affects the computational results of the RF full-wave simulations. To ensure acceptable accuracy, the voxel size should be at least one-tenth of the corresponding wavelength. As RF basic restrictions (SAR and APD) are volume-averaged values, staircasing artifacts associated with voxel-based human models have a less significant effect on dosimetry computations than those at low frequencies. Intercomparison studies have been conducted to validate computations using a complex voxel-based model for LF [43-46] and RF dosimetry problems [47-49].

## C. Exposure Assessment Standards: IEC TC106

The International Electrotechnical Commission (IEC) Technical Committee (TC) 106 was established in 1999 to develop international standards for measuring and calculating human exposure to electric, magnetic, and EMFs up to 300 GHz. Its scope includes characterizing the electromagnetic environment with regard to human exposure, defining measurement methods, instrumentation, procedures, and calculation methods. Notably, it excludes establishing exposure limits and mitigation methods.

In 2015, Working Group (WG) 9 was established under TC 106 to study exposure evaluation methods for WPT. Before that, no related standards had been developed. Initially, attention was focused on WPT technologies for mobile phone terminals and EVs.

Consequently, Technical Report (TR) 62905 was compiled and published in 2018 as an exposure evaluation method for WPT operating at frequencies up to 10 MHz [50]. Subsequently, the IEC/IEEE Joint Working Group (JWG) 63184 was established to develop international standards based on TR 62905, resulting in the issuance of IEC/IEEE 63184 in 2025 [9].

Regarding WPT using microwaves, TR 63377 [10], an exposure evaluation method for radio waves between 30 MHz and 300 GHz developed by WG9, was issued in 2022. Consequently, IEC/IEEE JWG 63480 has been established, and standardization efforts are ongoing.

### D. Overview of Relevant Product Standards

WPT systems often generate non-uniform, localized EMFs and involve scenarios in which the body is close to radiation sources. These conditions differ from the uniform field assumptions used to derive the reference levels. Consequently, directly comparing the measured spatial peak value of the field strength with the reference level may be overly conservative. Both the ICNIRP and IEEE state that exceeding the reference levels does not necessarily imply non-compliance, provided that induced fields or SAR remain within the basic restrictions [29]. To avoid over-conservatism, computational dosimetry using anatomical models and exposure-specific simulations (or measurements) is becoming increasingly necessary.

Although several general assessment standards exist, IEC/IEEE 63184 [9] and IEC TR 63377 [10] are particularly relevant to WPT exposure assessment methods and are hereby summarized. IEC/IEEE 63184 outlines "Assessment methods for human exposure to electric and magnetic fields from WPT systems—covering models, instrumentation, measurement, and computational methods and procedures (frequency range of 3 kHz–30 MHz)." It specifically defines multiple applicable assessment methods for stationary WPT systems near non- uniform magnetic fields.

Four tiers (1–4) of evaluation methods for direct effects (internal electric field, current density, or SAR) are prescribed in the IEC/IEEE 63184 [9]. Manufacturers can select any of these tiers:

1) Evaluation based on coil current of the WPT system (Tier 1): The maximum permissible coil current of the WPT system can be estimated using the reference level in the Biot–Savart law. The approximation of the formula defined in [9] is conservative for fields in the plane of the coil outside its circumference.

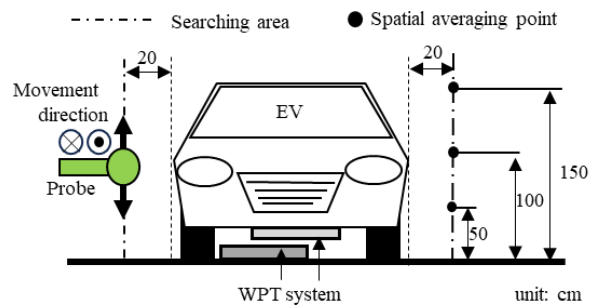

Fig. 2. Search area and spatial averaging points for EV use.

2) Evaluation of incident fields against reference levels (Tier 2): The peak electric and magnetic fields generated by the WPT system are identified in regions accessible to humans. Spatial averaging methods may be employed depending on the applied exposure guidelines (ICNIRP or IEEE). The evaluation area and spatial averaging points (Fig. 2) for the EV use case are prescribed in [9]. This assessment can be performed using either measurement or computational methods.

3) Evaluation of incident magnetic field strength using coupling factor (Tier 3): The incident field strength, derived by multiplying the measured or calculated magnetic field strength by the coupling factor, can be evaluated. The general concept of the coupling factor is defined in IEC 62311 [51]. The IEC/IEEE 63184 shows the coupling factor can be calculated in terms of localized exposure scenarios or characterized with respect to the normalized gradient of the incident magnetic field at the exposure evaluation location.

4) Evaluation of internal electric field, current density, and/or SAR against basic restriction (Tier 4): The internal electric field, current density, and SAR can be estimated through numerical simulation using anatomical human models for comparison with basic restrictions. SAR can also be measured if IEC/IEEE 62209-1528 [52] is applied.

As outlined in the four tiers of evaluation methods, the IEC/IEEE 63184 defines multiple approaches to avoid over conservatism because WPT systems below 30 MHz often generate non-uniform, localized EMFs and involve scenarios in which the body is near the radiation source.

IEC TR 63377 [10] outlines "procedures for the assessment of human exposure to EMFs from radiative WPT systems—measurement and computational methods in the frequency range of 30 MHz–300 GHz." Compared to WPT operating below 30 MHz, those above 30 MHz—called radiative WPT in the TR—allow for greater distances to be covered. Therefore, different evaluation methods are required based on exposure distance. IEC TR 63377 adapts the SAR evaluation methods for mobile phone terminals (IEC/ IEEE 62209-1528 [53]) and base stations (IEC 62232 [54]). For exposure distances within 20 cm of the WPT equipment, SAR measurements for whole-body and localized exposure can be performed using IEC 62232 and IEC/IEEE 62209-1528, respectively. For distances greater than 20 cm, whole-body SAR assessment based on IEC 62232 remains applicable.

If the frequency of the WPT exceeds 6 GHz, the incident power density is used instead of the SAR for the exposure assessment. IEC/IEEE 63195-1 [55], a method for measuring incident power density from mobile phones and similar devices, can be applied. Regarding the numerical calculation methods, IEC/IEEE 62704-1 [39] is used as the calculation method for local SAR, while IEC 62232 is used for whole-body averaged SAR.

MPT systems employ mitigation techniques, such as proximity sensors and time-period power control, to reduce RF-EMF exposure. Proximity sensors detect the presence of nearby people, triggering a reduction in radiated power. Additionally, the time-period power control reduces the maximum output power of the antenna after a fixed period. Therefore, TR 63377 describes a proximity sensor and a time-period power control verification method to ensure proper functionality of the technology in advance.

IEC/IEEE 63184 includes review-style discussions on exposure assessment methodologies. This standard concisely summarizes computational techniques and measurement approaches (e.g., [56-59]), serving as valuable reference sources and thus not reiterated in this review. Note that if the system does not comply with the IEC/IEEE 63184 standard, a mitigation strategy is required, as demonstrated in Section V.

### E. Electromagnetic Interference (EMI) with Medical Devices

With the increasing deployment of WPT systems in public and commercial environments, their potential electromagnetic effects on implantable medical devices (IMDs), such as pacemakers, have drawn significant attention from researchers and regulatory bodies owing to safety concerns.

Pacemakers are critical medical devices that regulate cardiac rhythm and provide life-saving support to patients with arrhythmia or other heart conditions. However, the functionality of pacemakers can be compromised by EMF exposure. Numerous studies have identified specific frequency ranges and intensities of EMFs that can interfere with pacemaker operations, causing malfunctions such as inappropriate pacing, inhibition, and mode switching [60-66]. Regulatory bodies, including ICNIRP and IEEE, have established guidelines [21, 22] for EMF exposure to minimize risks. However, real-world incidents highlight persistent vulnerabilities near devices such as electronic article surveillance (EAS) and WPT systems that generate strong magnetic fields in relatively LF bands where EMI effects can occur [67-71]. Furthermore, particularly in environments with high-intensity or unexpected EMF exposure—such as industrial zones, during medical procedures such as magnetic resonance imaging, or near devices generating strong magnetic fields—the potential impact of metal heating should be considered.

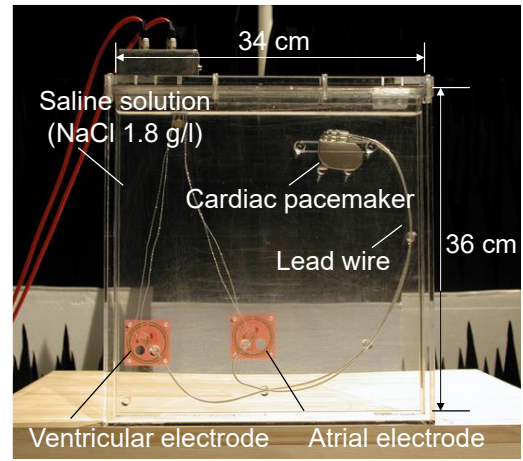

Fig. 3. Vertical torso phantom for pacemaker EMI testing.

Following international safety standards [72-78], commercially available pacemakers must be certified to comply with electromagnetic compatibility requirements as a condition for regulatory approval. However, the mechanisms underlying EMI in pacemakers are diverse and depend on the type of interfering source, including the conducted currents, time-varying magnetic fields, and high-voltage alternating electric fields [62]. These effects are influenced by field strength, frequency, waveform characteristics, and signal patterns. Moreover, as sensitivity and immunity specifications vary among pacemaker models, a comprehensive evaluation using multiple devices is essential to ensure sufficient market coverage.

To conduct EMI testing on pacemakers, it is necessary to replicate the conditions under which the device is implanted and sense intracardiac signals. Additionally, mechanisms for detecting EMI-induced malfunctions are essential. Human-body phantoms were used to simulate implantation conditions, with two main types: vertical and horizontal. For example, the vertical phantom, shown in Fig. 3, comprises an acrylic tank and is suitable for testing scenarios involving large stationary sources, such as EAS gates, RFID readers, and WPT coils, where EMF is nearly perpendicular to the torso. In contrast, the horizontal phantom is more appropriate for evaluating compact, portable sources, such as mobile phones and smartphones, used near the chest.

These phantoms are based on a modified version of Imich's model [61], incorporating a NaCl solution [77] and embedded electrodes for simultaneously injecting pseudo- electrocardiographic signals and detecting pacing pulses. Although pacemaker dimensions, shapes, and lead designs vary based on manufacturer and model, this testing system accommodates interchangeable units and leads, enabling consistent evaluation across different devices. Recently, emerging technologies, such as subcutaneous implantable cardioverter defibrillators and leadless pacemakers, which eliminate the need for transvenous or intracardiac leads, have been approved and are increasingly being adopted in clinical practice. The phantom system described here is adaptable for evaluating these next-generation devices.

It is worth mentioning that metallic implant safety in WPT is also evaluated (e.g., [79, 80]), although they are not scope of the standards.

## IV. REVIEW OF ASSESSMENT STUDIES

### A. Microwave Power Transmission

MPT systems can induce electromagnetic exposure in various scenarios, including unintended exposure of the wave path, use of wearable devices with MPT systems, and occupational exposure. Studies primarily focus on addressing these concerns through wave control and shielding methods and assessing radiation doses in specific scenarios to ensure compliance with international exposure guidelines.

In [81], the convex optimization approach enabled a more precise beam focusing in MPT at 900 MHz, thereby reducing unintended exposure to the surrounding areas. This targeted delivery can help minimize electromagnetic exposure to humans and enhance protection in practical deployments. In [82], the human protection issue in MTP at 5.8 GHz was addressed by formulating the safe charging problem, which ensures that exposure levels remain below a certain limit while maximizing energy delivery. Similar approaches were evaluated in other studies [83, 84]. In [85], human protection in MPT was achieved by designing antenna excitation patterns that steer beams away from human-occupied regions. By incorporating spatial constraints into beamforming optimization, the system achieves efficient power delivery while minimizing human exposure.

### B. Inductive Coupling-based WPT

Research on inductive coupling-based WPT has mainly been focused on enhancing transfer performance regarding the efficiency and power, but not so much on EMF protection. For instance, [86] presented a 2 kW inductive coupling WPT system that incorporated circular ferrimagnetic plates and metallic shielding. The results demonstrated that the system complies with the exposure regulations of the Australian Radiation Protection and Nuclear Safety Agency (based on the ICNIRP guidelines). In [87], the SAR and *in situ* electric field were assessed within an anatomical model exposed to an inductively coupled WPT system compliant with wireless power consortium coils at 140 kHz. The results showed that evaluating the chest region is critical for effective compliance with fundamental restrictions.

### C. Resonant Coupling-Based WPT

Extensive research has instead been conducted on dosimetric evaluations of magnetic resonant coupling-based WPT systems. In [88, 89], a 4-coil WPT system was developed to deliver 60 W of power using two coaxially aligned helical coils operating at 10 MHz. The measured electric and magnetic field values exceeded the reference levels prescribed by the exposure guidelines. In [90], the SAR in an anatomically-based human model was calculated for various exposure scenarios of a 10 MHz WPT system. The local 10 g averaged SAR limit was more restrictive than the whole-body averaged SAR limit. In [91], local and whole-body SAR limits were evaluated for low-MHz WPT systems, showing that local SAR found to vary by over 3 dB across different human anatomical models due to body-shape and tissue-property differences. In [92], 10 g averaged SAR was evaluated for 6.78 MHz resonant-type WPT systems with single and double receiving coils, showing that only marginal SAR increases with double receivers, lower SAR in child than adult models. In [93], the locations of peak exposure differed among the metrics, including the 99th percentile value of the electric field, average current density over 1 cm$^2$, and SAR for printed WPT system at 10 MHz. In [94], human exposure to EMFs from resonant WPT systems at 100 kHz, 6.78 MHz, and 13.56 MHz was analyzed with a 5-W transmission power. The results showed that SAR values remain below international guideline limits and that simplified models provide comparable accuracy to realistic models.

Various coil optimization strategies have been proposed to enhance power transfer efficiency [95-98], reducing leakage fields. In [97], at multi MHz, lightweight air-core coils integrated with resonant power converters enabled sufficient coupling to support battery-free drone operation and unrestricted hovering within the transmitter's proximal region. The validated SAR values were 1000 times below the exposure limits. In [98], a novel near-field plate was proposed to control leakage while maintaining high efficiency. A phantom that modeled human muscle tissue was positioned adjacent to the transmission loop. This configuration demonstrated a 93% reduction in computed SAR.

Room-scale transfer cavities have been proposed to enable WPT to multiple or moving devices across a large target area. In [99], a quasistatic cavity resonance was proposed for room-scale kilowatt-level WPT to mobile receivers at 1.3 MHz. Compliance assessment with SAR and electric field strength confirmed that 1.9 kW can be delivered at 90% efficiency. In [100], a system using the multimode quasistatic cavity resonance approach was demonstrated, achieving a power delivery efficiency exceeding 37.1% within a 3 m × 3 m × 2 m cavity at 1.2 and 1.34 MHz. This method showed the potential to deliver over 50 W of power to receivers while maintaining compliance with exposure guidelines.

### D. WPT for Electric Vehicles

Significant global efforts have been dedicated to developing conservative assessments for practical scenarios in EV WPT systems [9, 101]. Resonant coupling coils are used for this application, and the frequency prescribed in the standard is around 85 kHz [102].

In [103], a dosimetric study was conducted on three human body models positioned adjacent to a perfectly conductive vehicle at various human–vehicle angles. One notable observation was that the induced electric field in the child model was smaller than that in the adult model, which could be attributed to the reduced cross-sectional area of the child's body. This finding can be explained by Faraday's law, which states that the induction of an electric field is reduced in smaller-body models. Additionally, dosimetry has been conducted for models of pregnant women and children [104]. Various postures, such as lying on the ground with the right arm extended toward the WPT coils, were evaluated as worst-case scenarios in [105]. At a transfer power of 7 kW, the exposure of the hand beneath the vehicle remained well below the limits set by international guidelines, due to the smaller cross-sectional area of the hand compared to the torso. A similar computation was performed in [106] for a 7.7 kW WPT system, confirming nearly the same conclusion.

Furthermore, the reliability of numerical dosimetry itself has been validated in [107], where finite-difference time-domain (FDTD) simulations at 140 kHz demonstrated that coupled and co-simulation approaches yield results consistent with full-scale monolithic calculations. This provides additional confidence in the robustness of computational methods used for WPT exposure assessment. Some studies have evaluated the induced electric field in human models with metallic implants, suggesting an enhanced field strength [80, 108] (See also Sec. III E).

Modeling of the car body, both in terms of geometry and material, is another aspect that has been increasingly improved over the years. It started with a simple thin metal plate, as recommended in SAE J2954 [102], to mimic a vehicle floor pan [109, 110], followed by rough chassis geometries [103, 105, 111] and sophisticated CAD models of realistic cars [106, 112-114]. The latter study [114] highlighted the necessity to avoid unnecessary CAD details in order to maintain the computational efforts at a reasonable level and also pointed out the influence of the chassis geometry on the magnetic field distribution around and inside the car. The impact of the vehicle materials on the induced electric field was assessed in [111, 112]. In contrast to metallic vehicle bodies, carbon fiber reinforced plastic allows the magnetic field to penetrate the vehicle, leading to excessive exposure to the driver's seat. The induced electric field was pronounced around the buttocks, but still within exposure limits. This is not the case of high-power systems for heavy vehicles, such as minibuses, where the limits for vehicle occupants are sometimes exceeded, especially for misaligned charging conditions [115]. Instead, the impact of the body material was minimal for a human standing outside the vehicle. Although no dosimetry study is provided, the effect of coil parameters on the leaked magnetic field has also been investigated in [116]. Such findings complement the studies on vehicle geometry and materials by clarifying design factors that influence the external field environment.

Finally, a compliance assessment of a dynamic WPT system was recently investigated in [113]. Although a transfer power of 10 kW was accounted for, the induced electric fields remained below the limits provided by the ICNIRP for several human postures of passengers and bystanders and for several car materials of a compact vehicle.

Based on these studies, it is crucial to evaluate realistic postures in scenarios involving EV usage, particularly with respect to vehicle body materials and WPT technology (static vs. dynamic). Since the degree of overconservativeness between reference levels and basic restrictions is about 10 or more [117, 118], the induced electric field is more suitable to be evaluated for EV WPT systems. To address the issue of excessive conservatism in practical compliance assessments, a coupling factor outlined in the IEC standards can be used as an alternative to detailed numerical calculations. Although various definitions of the coupling factor have been presented [119, 120], they have not been elaborated in the context of EV WPT. Therefore, a simple assessment method based on the correlation between the external spatial averaged magnetic field at different points outside or inside the vehicle and the internal electric field has been proposed in [58, 59].

### E. WPT with Metasurface and Metamaterial

Various methods have been proposed to enhance system transfer performance. For WPT applications, metasurfaces or metamaterial slabs are commonly placed between or around transmitting and receiving coils to guide the magnetic fluxes toward the receiving coil, markedly enhancing power transfer efficiency and reducing magnetic field leakage and associated exposure levels [121-127]. In [124], a magnetically dispersive surface was proposed to enhance the power transfer efficiency of WPT systems at 5.7 MHz. It has been shown to enhance efficiency while generating a low peak-induced electric field, making it a promising solution for applications that require mitigation of SAR exposure concerns. In [127], a metasurface comprising a planar spiral resonator array positioned near the driver loop at 6 MHz has been shown to notably reduce the magnitude of the electric field by 60%, thereby decreasing SAR.

### F. WPT for Medical Devices

The global market for IMDs is expanding significantly, encompassing orthopedic implants, cardiovascular devices (such as pacemakers and defibrillators), neurostimulators, cochlear and ophthalmic implants, ingestible capsules, and microrobots. Among these, active IMDs (AIMDs), which are electric or electronic, are typically powered by batteries requiring surgical replacement. WPT offers a promising alternative, eliminating the need for physical connectors and enabling safer, longer-lasting implants [128, 129]. Although exposure guideline assessments are not directly applied to such applications, this section summarizes their features and assessment examples, which support safe use and efficiency improvements.

WPT for AIMDs is categorized by transmitter/receiver setups, operating frequency, field type (near-, mid-, or far-field), coupling method, power level, implantation depth, and whether implants are battery-powered or battery-free [129-131]. Although commercial adoption remains limited, several AIMD products employing WPT technology have already been developed. The advantages of WPT include: (1) reduced need for surgical interventions through external recharging, eliminating frequent battery replacements; (2) extended operational lifespan of implants through continuous or periodic wireless charging; (3) less invasive design enabled by smaller internal batteries and no transcutaneous wires or connectors; (4) enhanced patient comfort and mobility through discreet, user-friendly external power management; (5) improved safety, with reduced risks of infection and mechanical failure owing to the elimination of physical connectors; and (6) support for advanced functionality, as continuous power availability enables the use of energy-demanding features previously limited by battery constraints. High power requirements and miniaturization constraints complicate the design of effective deep-implant systems. Devices range from ultralow power (such as pacemakers, <100 μW) to high power (such as cardiac pumps, up to 25 W), with deeper implants being particularly challenging to power wirelessly. Despite these benefits, challenges such as tissue-induced power loss in deeper implants [132], EMI [68], and safety concerns continue to hinder practical deployment.

Scientific progress has been made in addressing these problems. For example, (1) left ventricular assist devices utilize a WPT system with a subcutaneous receiver and internal battery, reducing driveline infection risks and enabling short-term wireless operation [133]; (2) future AIMDs may share power through an internal wired network powered by a single WPT source, improving efficiency and reducing the need for multiple implants and surgeries [134].

Efforts to ensure compliance with exposure guidelines for implantable WPT systems have been reported considering implant-specific factors such as geometry, frequency selection, and coupling efficiency at 300 kHz and 13.56 MHz [135]. SAR is reduced by operating in the out-of-phase resonant mode, which lowers tissue electric fields, while biocompatible coatings further suppress localized SAR by isolating fields [136]. Optimized parasitic patches placed on the body surface improve power transfer efficiency while reducing SAR in 10 MHz band [137]. In addition, theoretical analyses have evaluated EMF exposure in human tissues to establish safe operating conditions for far-field RF-powered implantable sensors at 900 MHz, 2.4 GHz, and 5.8 GHz

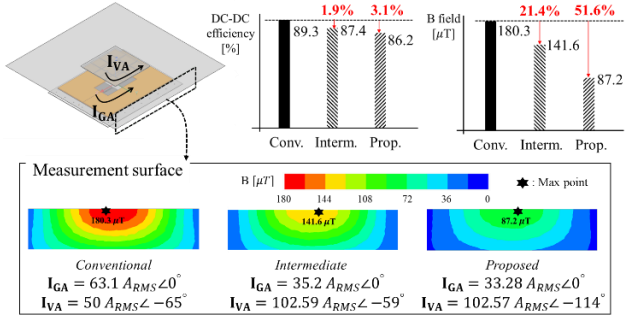

Fig. 4. EMF reduction method in an EV wireless charging system achieved by adjusting only the current magnitude and phase without using additional materials.

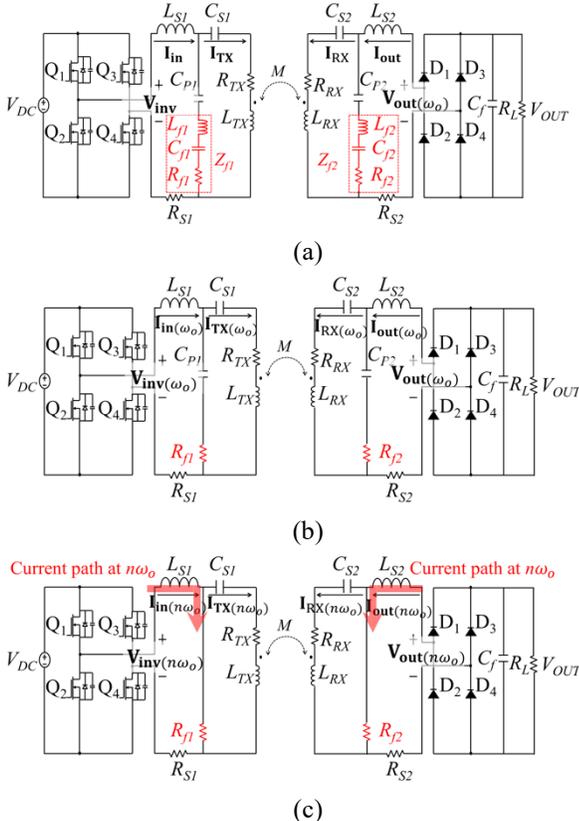

Fig. 5. Method for selectively mitigating radiated noise: (a) proposed LCC–LCC topology, (b) equivalent circuit at the operating frequency, and (c) equivalent circuit at harmonic frequencies.

[138]. A modified birdcage coil was proposed for an inductively coupled WPT system operating at 13.56 MHz. Compared to the conventional coil, SAR was reduced by improving magnetic field uniformity [139].

Ongoing research is crucial to overcome limitations related to efficiency, safety, and standardization. Although current environmental exposure limits are not directly applicable, exposure assessment remains essential for ensuring safety and efficiency.

## V. MITIGATION

As WPT technologies continue to expand across various applications from EVs to medical devices, concerns regarding EMF exposure and interference have gained increasing attention. Effective mitigation strategies must be developed and evaluated across various scenarios to ensure compliance with international guidelines.

To address magnetic field leakage in WPT systems, several shielding techniques have been widely applied across various product domains. These include the use of conductive materials, magnetic shielding using high-permeability materials [140-144], active shielding methods with additional coils and power sources [145-149], and reactive shielding employing resonant capacitors and secondary coils [150-155].

Magnetic materials, such as ferrite, with low magnetic reluctance, are commonly employed to guide magnetic flux along a desired path, thereby reducing field leakage [140-144]. Conductive materials, often integrated below the ferrite layer or used as part of the enclosure structure, help suppress leakage fields by inducing opposing eddy currents. As WPT system power levels increase, traditional ferrite materials face limitations owing to magnetic saturation and thermal issues.

To address these challenges, recent studies have explored nanocrystalline materials as an alternative magnetic shielding solution [156]. Although nanocrystalline cores typically exhibit lower magnetic permeability than ferrite, they offer a significantly higher saturation flux density, making them suitable for high-power WPT applications. Moreover, nanocrystalline materials have demonstrated excellent thermal stability and reduced core losses. Importantly, despite their lower permeability, they achieve stray magnetic field reduction comparable to that of ferrite in many practical configurations, further supporting their viability as effective shielding for next-generation WPT systems.

Beyond these conventional shielding approaches, several studies have proposed novel methods to mitigate magnetic field emissions in WPT systems [157-160]. A novel resonant circuit design perspective was introduced in [157-159]. While traditional resonant circuit designs primarily focus on maximizing power capacity and transfer efficiency, these studies emphasized minimizing magnetic field leakage as the main design objective. In [159], a method to reduce magnetic field leakage in EV WPT systems by controlling only the current magnitude and phase difference was proposed. As shown in Fig. 4, the method in [159] achieved a 51.6% reduction compared to conventional approaches. This shift represents a significant conceptual advancement in WPT system design.

Furthermore, in [160], a method for selectively suppressing radiated noise at harmonic frequencies was proposed by integrating an additional LC filter. Fig. 5 (a) shows the resonant topology proposed in the study, in which two additional components, $L_f$ and $C_f$, are incorporated into the conventional LCC–LCC topology.

At the operating frequency, as shown in Fig. 5 (b), $L_f$ and $C_f$ resonate and allow the circuit to function identically to the conventional LCC–LCC configuration. However, at specific harmonic frequencies, as depicted in Fig. 5 (c), the combination of $C_p - L_f - C_f$ resonates, directing harmonic currents through only the first and fourth loops. Consequently, the zero-harmonic current ideally flows through the transmitter and receiver coils.

In addition, an enhanced design methodology for reactive shielding coils was proposed [161], addressing factors overlooked in earlier studies [150-155]. Specifically, although earlier approaches ignored the reduction in the effective inductance of transmitter and receiver coils owing to the addition of shielding coils, a method for quantitatively evaluating the equivalent inductance was introduced [161]. Moreover, it presents a novel technique for selecting the shielding coil's resonant frequency, eliminating the need for a series compensation capacitor on the receiver side in an LCC–LCC system.

## VI. FUTURE DIRECTIONS AND SUMMARY

### A. Exposure Guidelines

Several articles and reviews provide a data gap for standardization and future directions of exposure guidelines. The ICNIRP summarized the knowledge gap in LF [34] and RF [162] guidelines. IEEE ICES summarized the report of a workshop held in February 2024 [163]. In addition, a significant difference between the ICNIRP and IEEE standards exists at the reference level over the intermediate frequency bands. Collaborative efforts would suppress this gap, ideally for harmonization.

One of the data gaps listed in [162] is that the core temperature rise for whole-body exposure should be assessed, which is relevant for setting a reference level for far-field exposure. This is relevant to the assessment of MPT exposure. As the GHz band is considered for such scenarios, dosimetry assessment using a high-resolution human body model is needed (e.g., [164, 165]).

One issue related to product compliance is the introduction of a local reference level in these guidelines and standards for RF. However, the ICNIRP LF guidelines [28] do not have such metrics, while body part exposure (especially for limb exposure) is mentioned in the IEEE C95.1. Discussion of partial body exposure to LF is limited [166]; thus, further efforts are required to resolve this issue.

In addition, an ongoing ICNIRP project addresses EMF and the environment, discussing that current exposure guidelines also provide adequate protection for animals and plants. The findings from this study may offer valuable insights into applications such as space solar power stations. To ensure consistency and public confidence, collaborative standardization efforts of different standardization bodies for human exposure and product safety will remain essential in addressing these gaps.

### B. Exposure Assessment

IEC/IEEE 63184 only prescribes assessment methods for stationary WPT systems as mentioned in Sec III D. IEC TC 69, which develops standards for electrical power and energy transfer systems for EVs, has recently discussed dynamic WPT systems operating below 30 MHz. It has published the IEC Publicly Available Specification (PAS) 61980-6, which covers specific power transfer requirements for the off-board side of magnetic field dynamic WPT systems for electric vehicles and electric road vehicles [167]. Therefore, it is important to develop assessment methods of human exposure to dynamic WPT systems in the next edition of the IEC/IEEE 63184.

As mentioned in Sec. III C, the IEC/IEEE JWG 63480 is developing an international standard of exposure assessment methods for radiative WPT based on IEC TR 63377. The main task of JWG 63480 is to standardize the validation methods for exposure-reduction techniques. As the transmission power of MPT increases in the future, it is possible that limits in exposure guidelines will be exceeded. This means that reduction techniques will be necessary. Therefore, it is important to develop validation methods to confirm whether these techniques work reliably prior to operating MPT systems. These future developments are expected to provide not only enhanced safety assurance, but also clearer compliance pathways for manufacturers seeking to implement higher-power or mobile WPT technologies.

### C. Pacemaker malfunction

During EMI testing, it is verified whether the sensing and pacing functions of the pacemaker, defined by the model, sensitivity setting, lead polarity, and operational mode, are affected by electromagnetic exposure. Because the test parameters span various device models, sensitivity settings, lead configurations, and operating modes, a complete evaluation requires a considerable amount of time. In addition, there is an increasing need for the international standardization of these test parameters, device configurations, and evaluation methods to ensure consistency and global applicability. Currently, research is being conducted not only on experimental investigations but also on the use of numerical simulations as a potential solution to these challenges.

*D. Complex Exposure Scenarios*

An integrated consideration of both product standards and exposure guidelines, including inputs from regulatory bodies, is essential. In certain scenarios, such as when a person extends an arm directly beneath an EV WPT system, the basic restrictions may be exceeded depending on the transmitted power level. To address such situations, foreign object detection systems have been considered potential countermeasures.

Following the ICNIRP 2010 guidelines [28], the instantaneous electric field should be used as a metric for frequencies below 100 kHz. However, owing to the nature of instantaneous values, high-performance monitoring systems are required to ensure compliance. However, minor exceedances may not necessarily cause adverse health effects, given the substantial reduction factors incorporated into the exposure limit derivation.

Therefore, a multifaceted approach involving collaboration with regulators and public health experts is necessary to ensure protection and practical implementation.

*E. Mitigations*

Early-stage research on magnetic field mitigation primarily focused on using magnetic materials [140-144] and conductive elements to guide or suppress stray fields, as well as employing active [145-149] and reactive shielding methods involving additional coils and resonant capacitors [150-155]. Recently, there has been growing interest in approaches that reduce EMF exposure and EMI not through external shielding but via innovative resonant circuit design strategies.

Rather than relying on a single mitigation technique, future WPT systems are expected to benefit from the synergy of traditional shielding methods and newly developed circuit-based approaches. As power levels in modern wireless charging applications increase, these integrated strategies hold strong potential for achieving efficient and effective solutions to stray magnetic field problems. Future mitigation strategies should aim to ensure regulatory compliance while minimizing system complexity and cost, thereby enabling scalable deployment across diverse application domains.

*F. Concluding Remarks*

This paper outlines key future directions for the safe and effective deployment of WPT technologies. The coordinated evolution of exposure guidelines, improved assessment standards for static and dynamic systems, and robust EMI evaluation frameworks for IMDs are critical. Addressing complex exposure scenarios and developing integrated mitigation strategies will ensure protection while supporting innovation. These efforts collectively pave the way for harmonized international standards and widespread societal adoption.